\documentstyle[12pt]{dubna98}
\begin{document}

\begin{center}
{\bf PIONIUM LIFETIME CORRECTIONS\\}
\vspace*{1cm}
H. SAZDJIAN\\
{\it Groupe de Physique Th\'eorique, Institut de Physique Nucl\'eaire,\\}
{\it Universit\'e Paris XI, F-91406 Orsay Cedex, France}
\end{center}

\vspace*{.5cm}
\begin{abstract}
{\small Pionium lifetime corrections are evaluated in the frameworks of
constrained Bethe-Salpeter equation and chiral perturbation theory.
Corrections of order $O(\alpha)$ are calculated with respect to the
conventional lowest-order formula, in which the strong interaction
amplitude has been calculated to two-loop order with charged pion masses.
The total correction is found to be of the order of $(6.1\pm 2.0)\%$.}
\end{abstract}

\vspace*{1cm}

\section{Introduction} \label{s1}

Our aim is to evaluate the sizable corrections to the nonrelativistic
formula of pionium lifetime\cite{dgbthuptbnnt}:
\begin{equation} \label{e1}
\frac{1}{\tau_0} = \Gamma_0 = \frac{16\pi}{9}\sqrt
{\frac{2\Delta m_{\pi}}
{m_{\pi^+}}} \frac{(a_0^0-a_0^2)^2}{m_{\pi^+}^2} |\psi_{+-}(0)|^2,\ \ \
\ \ \Delta m_{\pi}=m_{\pi^+}-m_{\pi^0}.
\end{equation}
Here, $a_0^I$ is the strong interaction (dimensionless) $S$-wave $\pi\pi$
scattering length in the isospin $I$ channel and $\psi_{+-}(0)$ is the ground
state wave function of the pionium at the origin (in $x$-space). The strong
interaction scattering lengths $a_0^0$ and $a_0^2$ have been evaluated in
the literature in the framework of chiral perturbation theory ($\chi$PT)
to two-loop order of the chiral effective lagrangian\cite{gl1,bcegs,kmsf}.
\par
Formula (\ref{e1}) has been obtained by treating the strong interaction
scattering amplitude at threshold as a first-order perturbation with respect
to the Coulomb potential between the charged pions. We shall evaluate in the
following the corrections of order $O(\alpha)$ ($\alpha$ being the fine 
structure constant) to formula (\ref{e1}). Most of the details of the
calculations can be found in Ref. \cite{js2}.
\par

\section{Bound state equation} \label{s2}

To deal with the bound state problem, we use the constraint theory approach
to the Bethe-Salpeter equation\cite{js1}, which consists in reducing
the latter, by means of a constraint applied to the external relative energy,
to a three-dimensional equation. The constraint that is chosen is the
following:
$C(P,p) \equiv  (p_1^2-p_2^2) - (m_1^2-m_2^2) \approx  0$,
where $p_1$ and $p_2$ are the external particle momenta and $m_1$ and
$m_2$ their physical masses. One then establishes a three-dimensional
eigenvalue equation that takes the form
\begin{equation} \label{e3}
\widetilde g_0^{-1} \Psi = -\widetilde V \Psi ,
\end{equation}
where $\widetilde g_0^{-1}$ is the wave equation operator defined for
two-spinless particle systems by
$\widetilde g_0=-1/\big(p_1^2-m_1^2+i\epsilon\big)\big|_{C}=
-1/\big(p_2^2-m_2^2+i\epsilon\big)\big|_{C}$,
the index $C$ denoting the use of the constraint.
$\widetilde V$ is the potential, related to the renormalized off-mass shell
scattering amplitude $T$ by a Lippmann-Schwinger type equation:
\begin{equation} \label{e5}
\widetilde V = \widetilde T + \widetilde V\widetilde g_0\widetilde T ,
\ \ \ \ \ \ \ \widetilde T = \frac{i}{2\sqrt{s}} T\Big|_C ,
\end{equation}
where $s=P^2=(p_1+p_2)^2$. The amplitude $\widetilde T$ contains the usual
Feynman diagrams, where the external particles are submitted to the
constraint $C$. The second term in the right-hand side of the first of
Eqs. (\ref{e5}) generates an iteration series, the diagrams of which are
called ``constraint diagrams'', where the integrations relative to
the factor $\widetilde g_0$ are three-dimensional, because of the presence
of constraint $C$. The constraint diagrams cancel, in the $s$-channel,
the singularities of the two-particle reducible diagrams of $\widetilde T$.
As long as perturbation theory is concerned,
Eq. (\ref{e3}) is equivalent in content to the exact Bethe-Salpeter
equation, with, however, a different arrangement of the perturbation
series. The interest of Eq. (\ref{e3}) is that it generates a
three-dimensional systematic perturbation theory with covariant 
propagators\cite{js1}.
\par
We use for our calculations the chiral effective lagrangian\cite{gl1}
in the $SU(2)\times SU(2)$ case. We shall be working with the pseudoscalar
densities defined as $P^a=i\overline q\gamma_5\tau^a q$ ($a$=1,2,3),
where $q$ are the quark fields and $\tau^a$ the Pauli matrices. 
The off-mass shell
scattering amplitude is defined as the amputated connected four-point
Green's function of the pseudoscalar densities, multiplied by the
corresponding wave function renormalization factors $2B_0F_0Z_a^{-1/2}(p_a)$,
where $F_0$ is the pion decay constant $F_{\pi}$ in the chiral limit
($F_{\pi}=92.4$ MeV) and $B_0$ the quark condensate parameter\cite{gl1}.
With the above definitions, the amplitude $T$ is renormalization group
invariant. Furthermore, Eqs. (\ref{e5}) imply that the constraint 
propagators $\widetilde g_{0,ab}$ are kinematic operators 
with physical masses and are not concerned with the renormalization factors.
\par
For the analysis of the decay of the pionium into two neutral pions, it is
convenient to use a coupled-channel formalism\cite{rw,js2}. Defining
wave functions, potentials and propagators in matrix form with the indices
$\{+-\}$ for the charged pion sector and $\{00\}$ for the neutral pion
sector, Eq. (\ref{e3}) is transformed into two coupled equations. Eliminating
the wave function $\Psi_{00}$ in terms of $\Psi_{+-}$ and neglecting
unimportant unitarity correction factors, one ends up with a single
wave equation for $\Psi_{+-}$:
\begin{equation} \label{e7}
-\widetilde g_{0,+-}^{-1}\Psi_{+-} = \bigg[V_{Coul.}+\overline V_{+-,+-}
-\frac{1}{2} V_{+-,00,}\widetilde g_{0,00}V_{00,+-}\bigg]\Psi_{+-}.
\end{equation}
$\overline V_{+-,+-}$ represents the potential relative to the
transition in the charged sector and from which the Coulomb potential
$V_{Coul.}=-m_{\pi^+}\alpha/r$ has been separated. $V_{00,+-}$ represents
the potential relative to the transition from the charged to the neutral
sector, etc.. In the constraint propagators $\widetilde g_{0,+-}$ and
$\widetilde g_{0,00}$ it is the physical masses of the charged and neutral
pions that appear, respectively. The nonrelativistic Coulomb potential
will be considered as the zeroth-order potential and the remaining
potentials in Eq. (\ref{e7}) will be treated as perturbations.
\par
At the pionium energy, the operator $\widetilde g_{0,00}$ in Eq.
(\ref{e7}) lies in the scattering region of the $\pi^0\pi^0$
sector and therefore contributes with imaginary values in $x$-space, thus
inducing an imaginary part to the pionium energy.
\par

\section{Separation of quark-photon interaction terms} \label{s3}

In first-order of perturbation theory, it is the third term of the right-hand
side of Eq. (\ref{e7}) that contributes to the imaginary part of the energy
and hence to the decay width. The simplest approximation at this stage
consists in taking for the potential $V_{00,+-}$ ($=V_{+-,00}$) the strong
interaction amplitude $\widetilde T_{00,+-}^{str.}$ at threshold (deviations
from the pionium energy to the charged pion threshold yielding corrections of 
order $\alpha^2$), calculated to two-loop order. Defining $P_0 = P_{0R}-
i\Gamma/2$, one recovers, at leading order in $\Delta m_{\pi}$, the 
nonrelativistic formula (\ref{e1}). Corrections of order $\alpha$ to 
$\Gamma_0$ arise from the presence of additional terms of electromagnetic
origin in $\overline V_{+-,+-}$ and $V_{00,+-}$, as well as from second-order
perturbation theory contributions of the potentials.
It is natural at this stage to isolate from the
rest the electromagnetic interactions coming from quark-photon interactions.
The reason for this separation is that electromagnetic interactions with the
explicit presence of the photon field lead, in bound state
problems, or near threshold, to infra-red singularities which necessitate
a detailed treatment. This is not the case of the quark-photon type
interactions which operate only through mass shift effects and lead
to regular expressions that can readily be continued from the bound
state energy domain to the $\pi^+\pi^-$ threshold. 
\par
The quark-photon interaction is represented in the chiral effective
lagrangian at order $e^2p^0$  \cite{u} by the term $e^2C\langle QUQU^
{\dagger}\rangle$, where $U$ is the chiral field, $Q$ the quark charge
matrix and $C$ an unknown constant. Using for the field $U$ the
representation $U = \sigma + i\mbox{\boldmath$\pi$}.\mbox{\boldmath
$\tau$}/F_0$, $\sigma = \sqrt{1-\mbox{\boldmath$\pi$}^2/F_0^2}$,
where \mbox{\boldmath$\tau$} are the Pauli matrices and 
\mbox{\boldmath$\pi$} the pion fields, one finds: 
\begin{equation} \label{e10}
e^2C\langle QUQU^{\dagger}\rangle = -\frac{2e^2C}{F_0^2}\pi^+\pi^-.
\end{equation}
This term induces for the charged pions the mass shift
$\big(\Delta m_{\pi}^2\big)_{q\gamma}\equiv
\big(m_{\pi^+}^2-m_{\pi^0}^2\big)_{q\gamma}= 2e^2C/F_0^2$ ,
which is nonvanishing in the chiral limit and is numerically close to
the physical pion mass shift\cite{dgmly,egpdr}. It has no effect
on the scattering amplitude in lowest order, but acts essentially
through insertions (together with counterterms of higher order) in the pion
loop propagators, where the charged pion masses are replaced by their 
(almost) physical masses. 
\par
We then split the scattering amplitude ${\cal M}$ ($=-iT$) into two parts:
\begin{equation} \label{e12}
{\cal M}={\cal M}^{str.+q\gamma}+{\cal M}^{em.},
\end{equation}
where ${\cal M}^{em.}$ contains all electromagnetic interaction terms
generated by explicit photon fields of the chiral effective lagrangian,
while ${\cal M}^{str.+q\gamma}$ contains the strong interaction terms 
together with the quark-photon interaction terms.
\par
Let us now neglect the amplitude ${\cal M}^{em.}$. It can be shown\cite{js2}
that the potential $V_{00,+-}$, calculated from Eq. (\ref{e5}),
where one takes also into account the contributions of the constraint
diagrams, is equal, to order $\alpha$, to the real part of the on-mass shell
scattering amplitude $\widetilde T_{00,+-}^{str.+q\gamma}$ taken at threshold:
\begin{equation} \label{e13}
V_{00,+-}^{str.+q\gamma}={\cal R}e\widetilde T_{00,+-}^{str.+q\gamma}
(p_1^2=p_2^2=m_{\pi^+}^2,p_3^2=p_4^2=m_{\pi^0}^2,s=4m_{\pi^+}^2,
t=u=-\Delta m_{\pi}^2),
\end{equation}
where $\Delta m_{\pi}^2=m_{\pi^+}^2-m_{\pi^0}^2$.
(An error in Ref. \cite{js2} concerning the mass-shell condition of the 
neutral pions has been corrected.)
\par

\section{Second-order perturbation theory} \label{s4}

At this order, it is the interference of the last two potential terms of
Eq. (\ref{e7}) that contributes to the imaginary part of the energy.
\par
The contribution of the discrete states of the pionium spectrum is found
to be: $\big(\Delta \Gamma\big)_{discr.} = (\alpha/3) (2a_0^0+a_0^2)
\Gamma_0$.
\par
The contribution of the continuous spectrum is ultra-violet
divergent, with a linear and a logarithmic divergences. In order to
avoid these divergences, it is necessary to associate with the previous
terms contributions coming from the constraint diagrams present in Eq.
(\ref{e5}). The constraint diagram with one pion loop has a linear divergence
that cancels the previous one. The three constraint diagrams with one
pion loop and one photon exchanged between the two internal pions have
logarithmic divergences the sum of which cancels the previous logarithmic
divergence. The sum of the contributions of the continuous spectrum and of
the constraint diagrams is then finite and one finds:
$\big(\Delta \Gamma\big)_{cont.} + \big(\Delta \Gamma\big)_{constr.}
= 1.1 \alpha (2a_0^0+a_0^2) \Gamma_0$.
\par
The total contribution of the discrete and  continuous states, together
with the constraint diagrams, is:
\begin{equation} \label{e15}
\big(\Delta \Gamma\big)_{str.} =  1.5\alpha (2a_0^0+a_0^2) \Gamma_0
= 0.004 \Gamma_0.
\end{equation}
\par

\section{Vacuum polarization}  \label{s5}

Vacuum polarization in positronium and hydrogen-like atoms provides
corrections of order $\alpha^3$. In the pionium problem, vacuum polarization,
because of the presence of electron loops and of the large ratio of the
reduced mass of the two-pion system to the electron mass, yields
corrections of order $\alpha$. This point was emphasized by Labelle and
Buckley\cite{lb}. Using the nonrelativistic QED effective field theory
formalism\cite{cl}, they have found: $\big(\Delta \Gamma\big)_{vac.\ pol.}
=0.43\alpha\Gamma_0$. 
\par
In the present formalism, vacuum polarization contributes through the
potential $\overline V_{+-,+-}$ with a local term\cite{lp}. It contributes
to the decay width through the interference term between 
$\overline V_{+-,+-}$ and the last term of Eq. (\ref{e7}) at the second
order of perturbation theory. The contribution of the discrete states is 
found to be: $\big(\Delta\Gamma\big)_{discr}=0.03\alpha\Gamma_0$, while that
of the continuous states, which is finite, is $\big(\Delta\Gamma\big)_
{cont.}=0.38\alpha\Gamma_0$. The total contribution is:
\begin{equation} \label{e16}
\big(\Delta\Gamma\big)_{vac.\ pol.}=0.41\alpha\Gamma_0=0.003\Gamma_0,
\end{equation}
in agreement with the previous estimate.
\par

\section{Electromagnetic radiative corrections} \label{s6}

These corrections arise from pion-photon interaction terms  contained in
${\cal M}^{em.}$ [Eq. (\ref{e12})]. The chiral effective lagrangian
for the $SU(3)\times SU(3)$ case including electromagnetism to order
$e^2p^2$ was presented by Urech\cite{u}. It was presented for the
$SU(2)\times SU(2)$ case by Meissner, M\"uller and Steininger\cite{mms}
and by Knecht and Urech\cite{ku}. We use for our subsequent
calculations the $SU(2)\times SU(2)$ lagrangian with the notations 
of Ref. \cite{mms}.
\par
The pion-photon radiative corrections to order $e^2p^2$ are represented by
the pion self-energy corrections and the vertex correction of the lowest
order scattering amplitude of the process $\pi^+\pi^-\rightarrow
\pi^0\pi^0$. The sum of these contributions to the off-mass shell
scattering amplitude ${\cal M}_{00,+-}$ at the pionium energy has a 
spurious $O(\alpha^0)$ term. The latter is cancelled, however, by the
contribution of the corresponding constraint diagram of Eq. (\ref{e5})
(with one photon exchanged between the charged pions). 
The corresponding total contribution is:
\begin{eqnarray} \label{e17}
\bigg[{\cal M}_{00,+-}^{em.}+{\cal M}_{00,+-}^{em.(constr.)}\bigg]_
{\pi\gamma} &=& \frac{(s-m_{\pi^0}^2)}{F_0^2}\bigg[-\frac{\alpha}{2\pi}-
\frac{3\alpha}{4\pi}\Big(\ln(\frac{m_{\pi}^2}{\mu^2})+1\Big) - 4e^2k_3^r
\nonumber \\
& &\ -\frac{2e^2}{9}(k_2^r+10k_{10}^r)\bigg] + 2e^2\frac{m_{\pi}^2}
{F_0^2}\bigg[2k_3^r+3(k_7^r+2k_8^r)\bigg],
\end{eqnarray}
where the $k^r$'s are the renormalized low energy constants appearing in
the counterterm lagrangian and $\mu$ is the renormalization mass, chosen
in the following for the numerical estimates equal to the $\rho$-meson
mass (770 MeV). We have also incorporated in the mass term of the tree
level amplitude the mass shift of the neutral pion.
For the numerical estimate of the corrections, we have adopted the values
of the $k^r$'s obtained by Baur and Urech\cite{bu} (in the Feynman gauge)
using a resonance model for the saturation of sum rules and assigned to them
conservative 100\% uncertainties. The numerical values that we use (after
conversion to the $SU(2)\times SU(2)$ case) are, in units of $10^{-3}$:
$k_2^r+10k_{10}^r=122.2$, $k_3^r=6.4$, $k_7^r+2k_8^r=0.8$. One finds for
the modification of the decay width:
\begin{equation} \label{e19}
(\Delta \Gamma)_{\pi\gamma}=\big(-0.0015\pm 0.0075\big)\Gamma_0,
\end{equation}
where the uncertainty comes from the $k^r$'s. 
\par

\section{Electromagnetic mass shift corrections} \label{s7}

These corrections are those contained in the amplitude ${\cal R}e {\cal M}
_{00,+-}^{str.+q\gamma}$ [Eqs. (\ref{e12})-(\ref{e13})]. They are 
induced by the term (\ref{e10}) and by its counterterms of the effective 
lagrangian and the effect of which is the generation of mass shifts in
the scattering amplitude. The corrections are evaluated with respect to 
the strong interaction scattering amplitude, in which all pions have the
same mass, chosen to be the charged pion mass\cite{gl1,bcegs,kmsf}.
\par
In the strong interaction case, the pion mass is essentially determined
in terms of the quark mass parameter $\hat m$ \cite{gl1} and the quark
condensate parameter $B_0$. In lowest order, one has $m_{\pi}^2=
2\hat mB_0$. To one-loop order, one has:
\begin{equation} \label{e20}
(m_{\pi}^2)^{str.}=2\hat mB_0+\frac{(2\hat mB_0)^2}{F_0^2}
\Big(2l_3^r+\frac{1}{32\pi^2}\ln(\frac{m_{\pi}^2}{\mu^2})\Big)
\equiv 2\hat mB,
\end{equation}
where $l_3^r$ is one of the low-energy constants of the effective 
lagrangian.
\par
According to the result (\ref{e10}), the neutral pion mass
remains unchanged when the $O(e^2p^0)$ term is introduced in the strong
interaction lagrangian. Therefore, to order $p^4$ of the strong interaction
lagrangian and to order $e^2p^0$ of the electromagnetic interaction
the mass appearing in the left-hand side of Eq. (\ref{e20}) can be
identified with the neutral pion mass. Higher-order effects
in the masses play only a corrective role and modify slightly this
equality, which can be represented in the approximate form as
$m_{\pi^0}^2=2\hat mB+O(e^2p^2)+O(p^6)\simeq 2\hat mB$.
\par
Let us now rewrite the amplitude ${\cal R}e {\cal M}
_{00,+-}^{str.+q\gamma}$ by specifying in detail its kinematic and
parametric conditions, as emerging from Eqs. (\ref{e13}) and (\ref{e20}):
\begin{equation} \label{e21}
{\cal R}e {\cal M}_{00,+-}^{str.+q\gamma}=
{\cal R}e {\cal  M}_{00,+-}^{str.+q\gamma}(s=4m_{\pi^+}^2,
p_1^2=p_2^2=m_{\pi^+}^2,p_3^2=p_4^2=m_{\pi^0}^2,2\hat mB\simeq m_{\pi^0}^2).
\end{equation}
On the other hand, the strong interaction amplitude is defined in the
following way:
\begin{equation} \label{e22}
{\cal R}e {\cal M}_{00,+-}^{str.}=
{\cal R}e {\cal  M}_{00,+-}^{str.}(s=4m_{\pi^+}^2,
p_1^2=p_2^2=m_{\pi^+}^2,p_3^2=p_4^2=m_{\pi^+}^2,2\hat mB=m_{\pi^+}^2).
\end{equation}
We have to calculate the difference between the two previous amplitudes.
The latter is the result of three effects: (i) The dynamical effect of the
quark-photon interaction term (\ref{e10}), which acts through insertions
in internal pion loop propagators. (ii) Shift of the neutral pion
momenta squared, $p_3^2$ and $p_4^2$, from their value $m_{\pi^+}^2$ of the
strong interaction case to their final value $m_{\pi^0}^2$. (iii) Shift
of the mass parameter $2\hat mB$ from $m_{\pi^+}^2$ (strong interaction case)
to $m_{\pi^0}^2$. As long as one considers linear effects in the pion mass
shift, these three effects can be calculated separately\cite{js2}. We shall
stick to this approximation in the following. One then obtains:
\begin{eqnarray} \label{e33}
\Big(\Delta{\cal R}e {\cal M}_{00,+-}\Big)_{m.\ shift}&=&  
{\cal R}e {\cal M}_{00,+-}^{str.+q\gamma}-
{\cal R}e {\cal M}_{00,+-}^{str.}\nonumber\\
&=&\frac{\Delta m_{\pi}^2}{F_0^2}-\frac{m_{\pi^0}^2\Delta m_{\pi}^2}
{16\pi^2F_0^4}\bigg[\frac{22}{3}\ln\big(\frac{m_{\pi^0}^2}{\mu^2}\big)-
\frac{23}{6}\bigg]\nonumber\\
& &\ +\frac{m_{\pi^0}^2\Delta m_{\pi}^2}{F_0^4}\big(8l_1^r-4l_3^r)
+2e^2\frac{m_{\pi^0}^2}{F_0^2}\bigg[\big(k_2^r-2k_4^r\big)+
\big(k_7^r-2k_8^r\big)\bigg].\nonumber  \\
& &
\end{eqnarray}
The $l^r$'s and $k^r$'s are the renormalized low energy constants of the
strong interaction and electromagnetic interaction lagrangians, respectively.
The logarithms that appear in the strong interaction amplitude,
associated with the $l^r$'s, are calculated with the charged pion mass.
The masses that appear in the corrective terms are taken equal to the 
neutral pion mass.
The $\Delta m_{\pi}^2$ factor of the lowest-order term represents the
physical pion mass squared shift, for the $O(e^2p^2)$ terms of the
neutral pion mass, coming from pion-photon and
quark-photon interactions, have been included in the tree level
amplitude. We neglect, in our numerical evaluations, isospin breaking 
effects coming from the quark masses, which are small\cite{gl2,ku}. We use,
as in Sec. \ref{s6}, the numerical values of the $k^r$'s
obtained in Ref. \cite{bu} (in the Feynman gauge). These are, in units of
$10^{-3}$: $k_2^r=5.4$, $k_4^r=6.2$, $k_7^r-2k_8^r=-2.3$, 
$k_7^r+k_{11}^r=-0.3$. The constants $l^r$ are obtained from Refs.
\cite{gl1,bcegs}; they are, in units of $10^{-3}$: $l_1^r=-5.4\pm 3.9$,
$l_3^r=0.8\pm 3.8$, $l_4^r=5.6\pm 5.7$. We also take $F_0=88$ MeV \cite{gl1}. 
\par
The correction for the decay width is:
\begin{equation} \label{e34}
\frac{\big(\Delta \Gamma\big)_{m.\ shift}}{\Gamma_0}=0.049\pm 
0.004\pm 0.001,
\end{equation}
where the first uncertainty comes from the $l^r$'s and the second one from
the $k^r$'s. The contribution to the above number of the tree level 
correction, represented by the first term of the right-hand side of Eq.
(\ref{e33}) and coming from the effect of type (iii), is 0.037. Therefore,
the $O(e^2p^2)$ terms contribute with an amount of 30\% with respect to the
tree level effect. 
\par
Knecht and Urech\cite{ku} have provided, from a direct evaluation of
the generating functional to one loop, the expression of the on-mass
shell $\pi^+\pi^-\rightarrow \pi^0\pi^0$ scattering amplitude in the
presence of electromagnetism. A finite result at threshold is obtained
after subtraction of the infra-red singularity of the photon.
The sum of our evaluations (\ref{e17}) and (\ref{e33}) agrees with the
result presented by the above authors, to linear effects in $\Delta m_{\pi}$.
\par

\section{${\bf O(e^2p^4)}$ effects\protect\footnote{The content of this section
has been added after the Workshop. 
I thank E. A. Kuraev, P. Labelle and A. G. Rusetsky for drawing my attention
to this question and for discussions.}} \label{s8}

$O(e^2p^4)$ effects come from diagrams having a pion loop with one
photon exchanged or having two pion loops with a mass shift insertion.
Most of these effects are expected to be small.
However, enhancements may appear due to the presence of infra-red
singularities. The latter generally arise from kinematic regions where
the loop momentum $k$ is small. Two cases are distinguished. In the first
case, the component $k_0$ is of the order of ${\bf k}^2/m_{\pi}$; in the
second case, $k_0$ is of the order of $|{\bf k}|$. The singularities of
the first case, which are the strongest ones, are generally cancelled by
those of the corresponding constraint diagrams. The singularities of the
second case are generally cancelled either by those of the crossed 
diagrams (if present) or by those of the self-energy diagrams. A typical
example of such cancellations was met in the vertex correction evaluation
[Sec. \ref{s6}], where the sum of the contributions of the vertex diagram 
and of the corresponding constraint and self-energy diagrams was found to
be free of singularities.
\par
By considering groups of diagrams of the above kinds, one arrives at the
conclusion that the sum of all $O(e^2p^4)$ diagrams is free of infra-red
singularities (at the pionium energy). However, in our treatment of 
the second-order perturbation theory problem [Sec. \ref{s4}], 
contributions of the constraint diagrams of the diagram with a photon
exchanged between the two loop-pions were already taken into account to
cancel the ultra-violet divergence of the continuous spectrum of the
intermediate states. Therefore, the infra-red contribution of the 
corresponding four-dimensional diagram should also be isolated. This 
diagram has an infra-red singularity of the first kind mentioned above
and has been evaluated in the literature \cite{bdbs}. Its contribution
to the pionium decay width, leaving aside its ultra-violet divergent part,
is:
\begin{equation} \label{e34a}
\big(\Delta \Gamma)_{O(e^2p^4)}=-\frac{\alpha}{3}(2a_0^0+a_0^2)
\Big(2\ln\alpha+3\ln2+21\zeta(3)/(2\pi^2)\Big)\Gamma_0=0.006\Gamma_0.
\end{equation}
\par
The remaining groups of diagrams, being infra-red regular, will have
orders of magnitude fixed by $\chi$PT. Let us consider the $O(e^2p^2)$
radiative correction of Sec. \ref{s6}. The corresponding relative
correction to the decay width, without the counterterms $k^r$, is of order
$\alpha$ [Eq. (\ref{e17})]. Pion loops have typical relative order of
magnitude of 5\% (compared to 1). The number of the $O(e^2p^4)$ groups of
diagram being of the order of 25 and assuming their signs are distributed at
random, one arrives at an estimate of 0.2\% for the corresponding relative
contribution to the decay with. A similar contribution, 0.2\%, should also
be expected from the low energy constants of the counterterm lagrangian.
Mass shift corrections due to
quark-photon interactions have produced a 30\% effect at the $O(e^2p^2)$
level relative to the tree level, with a 1\% global effect [Eq. (\ref{e34})
and comment following it]. Assuming that the corresponding $O(e^2p^4)$
effect (two-loop diagrams) is of relative order of 40\% with respect to the
$O(e^2p^2)$ effect (taking into account the increase in the number of
diagrams), one arrives at an estimate of 0.4\% for its contribution.The sum
of the previous contributions is then 0.8\%. The total contribution of the
$O(e^2p^4)$ diagrams is thus:
\begin{equation} \label{e34b}
\big(\Delta \Gamma)_{O(e^2p^4)}=\Big(0.006\pm 0.008\Big)\Gamma_0.
\end{equation}
\par

\section{Summary and concluding remarks} \label{s9}

The corrections to the nonrelativistic formula of the pionium decay width
can be represented in the following form:
\begin{equation} \label{e35}
\Gamma=\Gamma_0\sqrt{\Big(1-\frac{\Delta m_{\pi}}{2m_{\pi^+}}\Big)}
\Big(1+\frac{\Delta\Gamma}{\Gamma_0}\Big),
\end{equation}
where $\Delta \Gamma$ is composed of the following contributions:
\begin{equation} \label{e36}
\Delta \Gamma=\big(\Delta \Gamma\big)_{str.}+\big(\Delta \Gamma\big)_
{vac.\ pol.}+\big(\Delta \Gamma\big)_{\pi\gamma}+
\big(\Delta \Gamma\big)_{m.\ shift}+\big(\Delta \Gamma\big)_{O(e^2p^4)}.
\end{equation}
One finds for $\Delta \Gamma/\Gamma_0$:
\begin{equation} \label{e37}
\frac{\Delta \Gamma}{\Gamma_0}=0.061\pm 0.004\pm 0.008\pm 0.008,
\end{equation}
the first uncertainty coming from the low energy constants $l_i^r$, the
second one from the $k_i^r$'s and the third one from the $O(e^2p^4)$
diagrams. $\Gamma_0$ [Eq. (\ref{e1})] is calculated with
the strong interaction scattering lengths evaluated up to two-loop order
($a_0^0-a_0^2=0.258$ \cite{bcegs}). One finds $\tau_0=3.19\times
10^{-15}$ s. With the corrective terms, the lifetime becomes:
$\tau=(3.02\pm 0.07)\times 10^{-15}$ s.
\par
The different contributions are summarized in Table 1.
\par
\begin{table}[ht]
\begin{center}
\begin{tabular}{|c|c|}
\hline
Effect & ${\Delta \Gamma}/{\Gamma_0}$ (\%)\\
\hline
strong & 0.4 \\
vac. pol. & 0.3 \\
$\pi\gamma$ rad. corr. & $-0.1\pm 0.7$ \\
mass shift & $4.9\pm 0.4 \pm 0.1$ \\
$O(e^2p^4)$ & $0.6\pm 0.8$ \\
\hline
total & $6.1\pm 2.0$ \\
\hline
\end{tabular}
\end{center}
\caption{The various corrections to the decay width.}  
\end{table}
\par
Evaluations similar to those of the present work were done by Ivanov {\it
et al.}\cite{illr}, solving the Bethe-Salpeter equation in the Coulomb
gauge and using, for the mass shift corrections, the result provided in
Ref. \cite{ku}. The results found by Ivanov {\it et al.} and ours are
qualitatively and quantitavely compatible among themselves.
In particular, the sum of their evaluations corresponding to strong and
one Coulomb exchange corrections ($-0.22+1.55$) should be compared with
the sum of our evaluations of strong and $O(e^2p^4)$ effects ($0.4+0.6$).
The latter effect, not evaluated in Ref. \cite{js2}, has reintroduced an
infra-red logarithm present in Ref. \cite{illr}. The sum of the quantities 
corresponding to mass shift and electromagnetic radiative corrections of Ref.
\cite{illr} ($2.99+1.73$) should be compared with the sum of our evaluations
of mass shift and pion-photon radiative corrections ($4.9-0.1$).
\par

\vspace*{1 cm}

\noindent
{\bf Acknowledgments}
\par
I would like to thank the organizers of the Workshop for having created a
stimulating atmosphere and for their hospitality.
\par


\vspace*{.5cm}

\begin{thebibliography}{50}

\bibitem{dgbthuptbnnt}S. Deser, M. L. Goldberger, K. Baumann and W. Thirring,
{\it Phys. Rev.} {\bf 96} (1954) 774;
J. L. Uretsky and T. R. Palfrey, Jr., {\it Phys. Rev.} {\bf 121}
(1961) 1798;
T. L. Trueman, {\it Nucl. Phys.} {\bf 26} (1961) 57;
S. M. Bilen'kii, Nguyen Van Hieu, L. L. Nemenov and
F. G. Tkebuchava, {\it Sov. J. Nucl. Phys.} {\bf 10} (1969) 469.
\bibitem{gl1}J. Gasser and H. Leutwyler, {\it Ann. Phys. (N.Y.)} {\bf 158}
(1984) 142.
\bibitem{bcegs}J. Bijnens, G. Colangelo, G. Ecker, J. Gasser and M. Sainio,
{\it Phys. Lett.} {\bf B374} (1996) 210; {\it Nucl. Phys.} {\bf B508} 
(1997) 263.
\bibitem{kmsf}M. Knecht, B. Moussallam, J. Stern and N. H. Fuchs,
{\it Nucl. Phys.} {\bf B457} (1995) 513; {\bf B471} (1996) 445.
\bibitem{js2}H. Jallouli and H. Sazdjian, {\it Phys. Rev.} {\bf D58}
(1998) 014011 and (E).
\bibitem{js1}H. Jallouli and H. Sazdjian, {\it Ann. Phys. (N.Y.)} {\bf 253}
(1997) 376.
\bibitem{rw}G. Rasche and W. S. Woolcock, {\it Nucl. Phys.} {\bf A381}
(1982) 405.  
\bibitem{u}R. Urech, {\it Nucl. Phys.} {\bf B433} (1995) 234.
\bibitem{dgmly}T. Das, G. S. Guralnik, V. S. Mathur, F. E. Low and
J. E. Young, {\it Phys. Rev. Lett.} {\bf 18} (1967) 759.
\bibitem{egpdr}G. Ecker, J. Gasser, A. Pich and E. de Rafael, {\it Nucl.
Phys.} {\bf B321} (1989) 311.
\bibitem{lb}P. Labelle and K. Buckley, preprint hep-ph/9804201.
\bibitem{cl}W. E. Caswell and G. P. Lepage, {\it Phys. Lett.} {\bf B167}
(1985) 437.
\bibitem{lp}E. M. Lifshitz and L. P. Pitaevskii, {\it Relativistic
Quantum Theory} (Pergamon Press, Oxford, 1974), p. 421.    
\bibitem{mms}Ulf-G. Meissner, G. M\"uller and S. Steininger, 
{\it Phys. Lett.} {\bf B406} (1997) 154; {\bf B407} (1997) 454(E).
\bibitem{ku}M. Knecht and R. Urech, {\it Nucl. Phys.} {\bf B519} (1998)
329.  
\bibitem{bu}R. Baur and R. Urech, {\it Phys. Rev.} {\bf D53} (1996) 6552;
{\it Nucl. Phys.} {\bf B499} (1997) 319.
\bibitem{gl2}J. Gasser and H. Leutwyler, {\it Nucl. Phys.} {\bf B250} (1985)
465.
\bibitem{bdbs}D. J. Broadhurst, {\it Phys. Lett.} {\bf B101} (1981) 423;
A. Djouadi, {\it Nuovo Cim.} {\bf 100A} (1988) 357;
M. Beneke and V. A. Smirnov, {\it Nucl. Phys.} {\bf B522} (1998) 321.  
\bibitem{illr}M. A. Ivanov, V. E. Lyubovitskij, E. Z. Lipartia and
A. G. Rusetsky, preprint hep-ph/9805356; V. E. Lyubovitskij, E. Z. Lipartia
and A. G. Rusetsky, {\it JETP Lett.} {\bf 66} (1997) 747.

\end{thebibliography}
\end{document}